\documentclass[prl,twocolumn,amsmath,amssymb,floatfix,letter]{revtex4}
\usepackage[dvips]{graphicx}
\usepackage{color,array,dcolumn}
\usepackage{amsmath}
\usepackage{amssymb}
\usepackage{xcolor}

\begin{document}

\title{Superfluidity meets the solid-state:\\ 
frictionless mass-transport through a (5,5) carbon-nanotube}


\author{Alberto Ambrosetti$^1$, Pier Luigi Silvestrelli$^1$ and Luca Salasnich$^{1,2}$}

\affiliation{$^1$Dipartimento di Fisica e Astronomia, Universit\`{a} degli Studi di Padova, via Marzolo 8, \textsl{35131}, Padova, Italy 
\\
$^2$ Istituto Nazionale di Fisica Nucleare, Sezione di Padova, via Marzolo 8, 35131 Padova, Italy,  and 
Istituto Nazionale di Ottica del Consiglio Nazionale delle Ricerche, Unit\`a di Sesto Fiorentino,
via Carrara 1, 50019 Sesto Fiorentino (Firenze), Italy
}

\begin{abstract}
\date{\today}
	Superfluidity is a well-characterized quantum phenomenon which entails frictionless-motion of mesoscopic particles through a 
superfluid, such as $^4$He or dilute atomic-gases at very low temperatures. As shown by Landau, the incompatibility between 
energy- and momentum-conservation, which ultimately stems from the spectrum of the elementary excitations of the superfluid, 
forbids quantum-scattering between the superfluid and the moving mesoscopic particle, below a critical speed-threshold.
Here we predict that frictionless-motion can also occur in the absence of a standard superfluid, i.e. when a He atom travels 
through a narrow (5,5) carbon-nanotube (CNT).
Due to the quasi-linear dispersion of the plasmon and phonon modes that could interact with He, the (5,5) 
CNT embodies a solid-state analog of the superfluid, thereby enabling straightforward transfer of Landau's criterion of superfluidity. 
As a result, Landau's equations acquire broader generality, and may be applicable to other nanoscale friction phenomena, whose 
	description has been so far purely classical.
\end{abstract}

\maketitle


Superfluidity~\cite{superf1,superf2,superf} is a well-characterized physical phenomenon, which enables frictionless-flow of a mesoscopic 
particle through a superfluid medium, such as $^4$He or dilute atomic-gases at very low temperatures. 
When the elementary excitations of the superfluid exhibit a quasi-linear spectrum, two-fold conservation of energy 
and momentum interdicts quantum-mechanical scattering, as long as the mesoscopic particle does not exceed a critical 
velocity-threshold. 
Below the critical velocity, the quasi free-particle spectrum of the mesoscopic body -which is quadratic 
in momentum- is incompatible with the spectrum of the superfluid, which is instead quasi-linear at small momenta. 
Most notably, while spectral-incompatibility is pivotal to superfluidity, Landau's theory~\cite{landau} does not 
invoke particular assumptions about the nature of the medium, as long as a {\it free} particle can pass through.
It is thus conceivable that seemingly disparate systems may eventually lead to analogous frictionless flow, 
implying non-trivial transferability of Landau's criterion of superfluidity.
Known extensions of the standard mechanism contemplate for instance supersolidity~\cite{chester,tanzi}, or even 
exciton-condensation~\cite{lozovik,depalo,high,perali} in two-dimensional solid nanostructures. 
However, one could question about the existence of {\it generalized-superfluid} mechanisms even in the normal state 
--i.e. in the absence of Bose-Einstein condensation--, as long as the essential requirements are met.

\begin{figure}
\includegraphics[width=8.5cm]{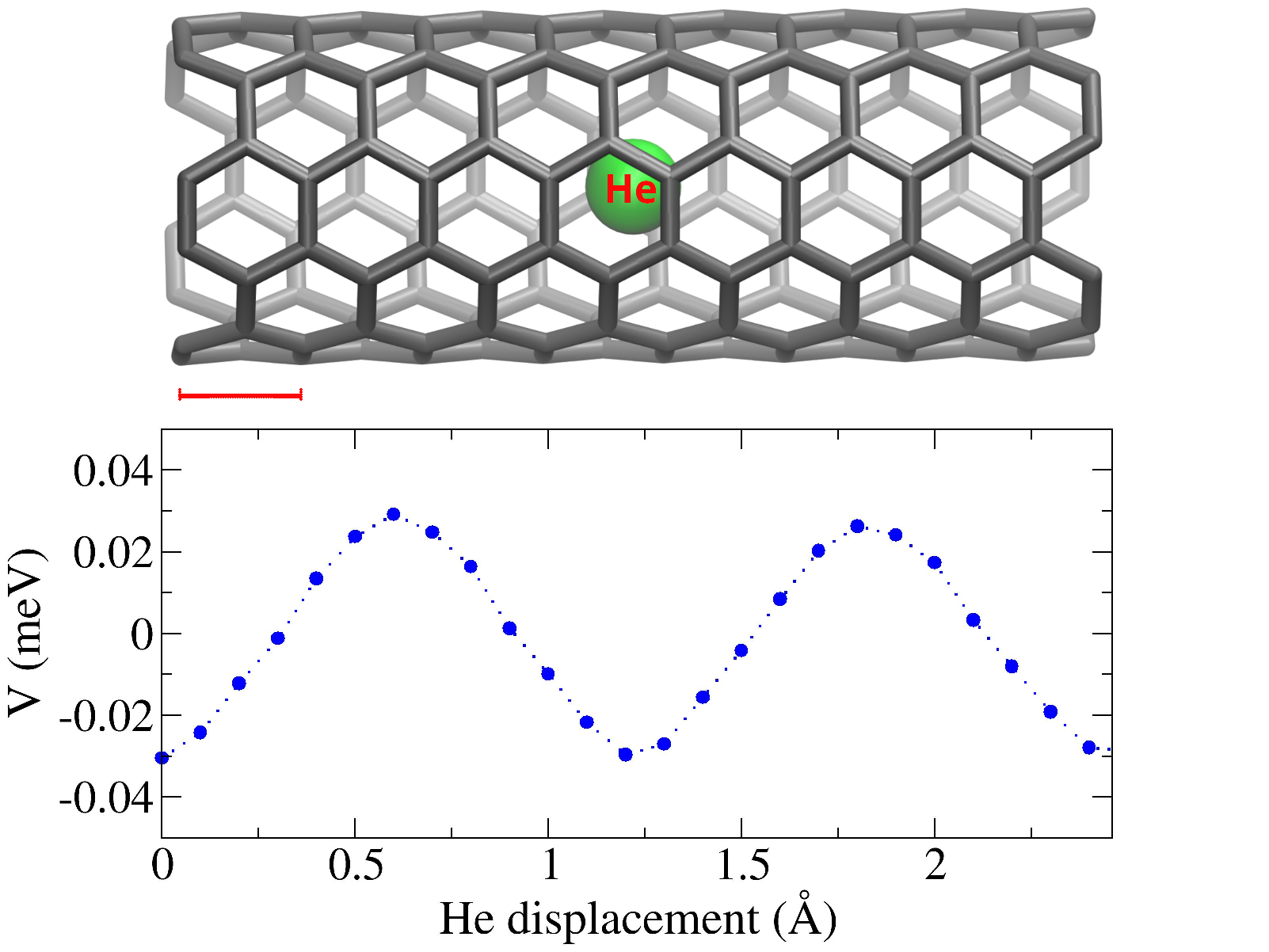}
	\caption{Potential $V_{\rm He}$ computed as a function of $x_{\rm He}$ (He displacement) within the (5,5) CNT unit cell. C atoms are fixed in the equilibrium position, 
	in the absence of electronic displacements. The geometry of the He atom confined in the (5,5) CNT $8\times1\times1$ supercell is illustrated for reference in the upper panel. The red segment visually indicates the longitudinal size ($L$) of the CNT unit cell.
	}
\label{fig1}
\end{figure}

To prove this idea true, in this Letter we consider a $^4$He atom moving through a (5,5) carbon-nanotube~\cite{dresselhaus} (CNT) 
-- which can be regarded as a closed, cylindrical-shaped  graphene~\cite{geim} tube (see Fig.~\ref{fig1}), 
characterized by a radius of 3.41 \AA\, and longitudinal metallicity. 
Scattering rates will be derived from scratch, without relying on assumptions
adopted in standard superfluidity (no ultracold gas is demanded).
A single He atom can fit in the center of the (5,5) CNT section, and it can easily move along the longitudinal axis. 
The dispersion of the relevant low-energy quasiparticles of the CNT, i.e. plasmon and phonon
excitations, bears formal analogies with the quasi-linear Bogoliubov's spectrum, so that the (5,5) CNT could act as an 
{\it effective superfluid medium}, providing on equal footing a viable channel for He transport.
We note in passing that low-dimensional nanostructures readily attracted scientific interest in relation to superfluidity~\cite{reatto, ancilotto},
although the presence of actual ultracold gases was so far always invoked. On the other hand, the evidence of 
ballistic electron transport~\cite{ballistic, ballistic2} in CNT's, adds even more appeal to these systems, also in view of their
availability as C allotropes with outstanding mechanical-resistance and chemical-inertness.

Hereafter we build a quantum mechanical model, based on first-principle density functional theory (DFT) simulations, 
relying on semi-local~\cite{PBE} exchange-correlation, and including dispersion~\cite{science, exact} corrections within Grimme's D2~\cite{d2} prescription. The approximations adopted are listed and discussed in detail in the
Supplementary Material \cite{SM}. 
The Quantum Espresso~\cite{qe} simulation package is exploited, 
in combination with ultrasoft pseudopotentials and an energy cutoff of 35 Ry for the plane-wave expansion of the electronic wavefunctions.
Since we are primarily interested in the flow of a single He atom, periodic DFT simulations will
minimize the interaction with periodic replicas by adoption of a long supercell (8 unit cell replicas along the CNT axis,
with a total length of 19.7\AA) and setting the transversal cell size to 15 \AA.
The Hamiltonian describing the one-dimensional (1D) motion of a He monomer along the CNT axis (indicated as $\hat{x}$ -- 
atomic units are adopted hereafter) is:
\begin{equation}
H_{\rm He}= -\frac{\partial_{x_{\rm He}}^2}{2 m_{\rm He}} + V_{\rm He}(x_{\rm He},\bf{R_{\rm ion}},\delta\pmb{\rho}_{\rm el})\,,
\label{ham-he}
\end{equation}
where $\bf{R_{\rm ion}}$ are the ionic coordinates, and $\delta\pmb{\rho}_{\rm el}$ the electronic charge displacements in the CNT.
To investigate the problem within a perturbative framework, we initially assume that all C atoms are fixed in the equilibrium configuration, and that no
electronic displacement takes place. In physical terms, this corresponds to the Born-Oppenheimer (BO) approximation, in combination with the electronic groundstate.
Under these assumptions, the potential energy $V_{\rm He}$ experienced by a single He molecule traveling along the (5,5) CNT is computed by DFT.
Transversal ($\hat{y}-\hat{z}$) He motion can be approximated by a 2D quantum harmonic oscillator model, whose frequency (estimated by DFT) 
$\omega_{\rm  He}$ is $\sim8$ meV.
As from Fig.~\ref{fig1}, $V_{\rm He}$ is a sinusoidal function of $x_{\rm He}$ (phases can be absorbed by rigid translation):
$V_{\rm He}(x_{\rm He})\sim V \sin(2 Q x_{\rm He})$, where $Q=2\pi/L$, and $L$ is the CNT unit cell
length (i.e. 2.46 \AA). The magnitude of the oscillations amounts to $V\sim 0.035$ meV {(and undergoes limited 
change when exact exchange is included - see \cite{SM})}. Dispersion 
interactions contribute to $V$ with only $\sim 3\times10^{-4}$ meV and can thus be neglected.
Notably, $V$ is about 200 times smaller than $\omega_{\rm  He}$; one accordingly expects that for sufficiently slow He, transversal 
excitations can be factorized from longitudinal translations. The motion of He is thus effectively reduced to 1D.

\begin{figure}
\includegraphics[width=8.8cm]{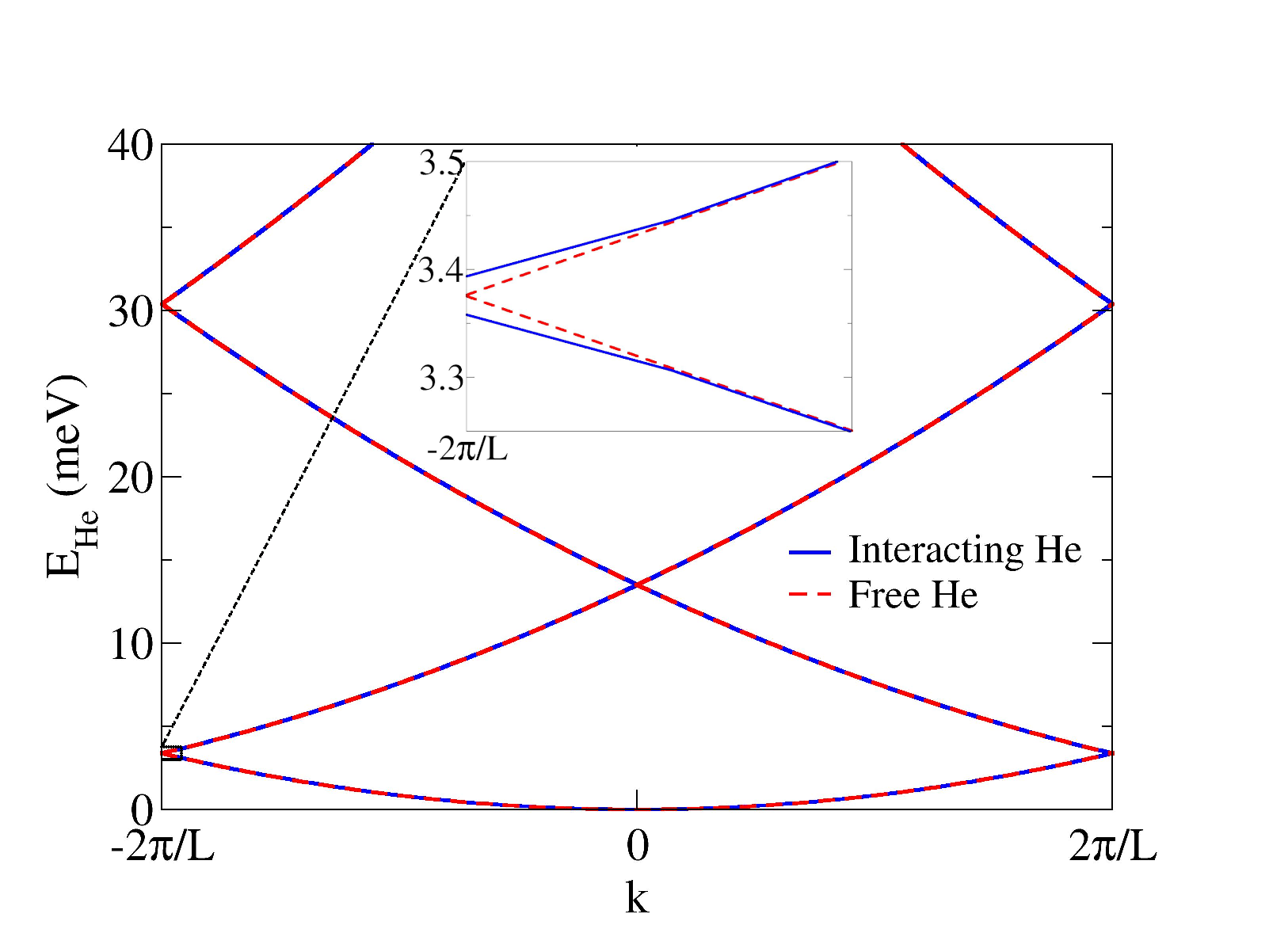}
	\caption{Spectrum of a single He atom (longitudinal modes), subject to the potential $V_{\rm He}$, as reported in Fig.~\ref{fig1}. Since the periodicity of $V_{\rm He}$ corresponds to half unit cell (i.e. $L/2$), the Size of the first Brillouin zone is rescaled correspondingly. Inset: detail of a band edge, where  $V_{\rm He}$ induces small splitting.}
\label{fig2}
\end{figure}

The He spectrum relative to longitudinal motion is obtained diagonalizing the Hamiltonian Eq.~\eqref{ham-he}, and it is hardly distinguishable from the free-particle 
dispersion $E_{free}(k)=k^2/(2 m_{\rm He})$, where $k$ is the (1D) $\hat{x}$ momentum (parallel to the CNT longitudinal axis). 
As from Fig.~\ref{fig2}, largest deviations are found at the Brillouin-zone (BZ) edges, where small band-splittings 
emerge due to the He-CNT coupling $V_{\rm He}$.
Due to the similarity between the free and interacting He spectra, one can estimate the atomic velocity as $v_{\rm He}\simeq q/m_{\rm He}$, while 
single plane-waves provide a good approximation for He eigenstates.

Both He-phonon and He-plasmon couplings provide possible channels for the He-CNT scattering. Clearly, 
the spectra of the available (phononic and plasmonic) excitations play a major role in this respect, eventually
determining the admitted transitions.
Given the small energy scales observed so far, one can assume that the lowest-frequency modes will be most
relevant.
Concerning phonon excitations, the (5,5) CNT is characterized by four acoustic modes~\cite{dresselhaus,zhang} whose frequency vanishes in the
$q\rightarrow 0$ limit, according to a linear dispersion $\omega_j(q)=v_j |q|$. The index $j$ runs over two degenerate transverse-acoustic modes (TA) with
$v_{\rm TA}=4.5\times10^{-3}a.u.$, a {\it twist} mode (TW) with $v_{\rm TW}=6.9\times10^{-3}a.u.$, and longitudinal-acoustic mode 
(LA) with $v_{\rm LA}=9.7\times10^{-3}a.u.$, according to existing~\cite{zhang,dresselhaus} literature.
A longitudinal plasmon with vanishing frequency at $q\rightarrow 0$ also exists, while transversal plasmon modes are gapped. 
Given the longitudinal metallicity of the (5,5) CNT, nearly-1D plasmons exhibit~\cite{Dobson} 
quasi-linear dispersion (up to logarithmic corrections - see \cite{SM}) 
in the long wavelength limit, and a tight-binding 
approach (see full derivation in Ref.~\cite{mbdc}) predicts typical plasmon velocities of the order of $\sim$1 a.u.

To address now the friction mechanism experienced by He in the CNT we consider that a travelling He atom can scatter against the CNT, 
transferring part of its energy either to CNT phonon modes or to plasmons. When scattering takes place, the kinetic energy
of the He atom decreases and this is traduced into an affective friction force. 
To describe this process, one needs to explicitly treat  
both ionic ($\bf{R_{\rm ion}}$) and electronic ($\delta\pmb{\rho}_{\rm el}$) degrees of freedom, overcoming the
BO approximation.
One accordingly expands the potential energy $V_{\rm He}$ to first order in both fluctuations, 
starting from equilibrium geometry ($\mathbf{\bar{R}_{\rm ion}}$) and zero charge displacements ($\delta\pmb{\rho}_{\rm el}=0$).
Ionic displacements are defined as
$\mathbf{\delta R_{\rm ion}}=\mathbf{R_{\rm ion}}-\mathbf{\bar{R}_{\rm ion}}$, and the expanded potential reads: 
\begin{eqnarray}
	V_{\rm He}(x_{\rm He},\mathbf{R_{\rm ion}},\delta\pmb{\rho}_{\rm el})=V_{\rm He}(x_{\rm He},\mathbf{\bar{R}_{\rm ion}},\delta\pmb{\rho}_{\rm el}=0)+ \nonumber \qquad  \\
	\sum_i \partial_{R_{i,\rm ion}} V_{\rm He}(x_{\rm He},\mathbf{R_{\rm ion}},\delta\pmb{\rho}_{\rm el}=0)|_{\mathbf{R_{\rm ion}}=\mathbf{\bar{R}_{\rm ion}}} \, \delta R_{i,\rm ion}+ \nonumber  \\ 
	\sum_i \partial_{\delta\rho_{{\rm el},i}} V_{\rm He}(x_{\rm He},\mathbf{\bar{R}_{\rm ion}},\delta\pmb{\rho}_{\rm el})|_{\delta\pmb{\rho}_{\rm el}=0}\, \delta \rho_{{\rm el},i}
+ ... \, \, . \qquad 
\label{ephcoupl}
\end{eqnarray}
This expression can be physically interpreted noting that the derivatives of $V_{\rm He}$ with respect to the ionic coordinates relate to ionic forces: 
$\partial_{R_{i,\rm ion}} V_{\rm He}= -F_{i,\rm ion}$. 
Instead, the derivative with respect to the $i$-th charge $\partial_{\rho_{{\rm el},i}}$ corresponds to an {\it effective potential} $\tilde{v}_i$ 
acting on site $i$.
Eq.~\eqref{ephcoupl} can thus be expressed in compact 
form as $-F_{i, \rm ion} \delta R_{i,\rm ion} + \tilde{v}_i \delta \rho_{{\rm el},i}$. Repeated indices are contracted for compactness and  
the same notation will be adopted  hereafter.
In Eq.~\eqref{ephcoupl} ionic and electron-charge motions are treated as 3D, at variance with He, and they naturally account for the quasi-1D geometry of the CNT. 

Both ionic and electronic-charge displacements are connected to quantum mechanical excitation modes of the CNT, i.e. phonons and plasmons.
In the case of phonons, there exists a unitary transformation that allows to express the geometry of the $j$-th collective vibrational modes with
(1D) wavenumber $q$ (i.e. $\delta \tilde{R}_j(q)$) in terms of the ionic coordinates. The transformation is  
$\delta \tilde{R}_j(q)=(1/\sqrt{N}) S_{j,n}^{\dagger} e^{iq l_c L} \delta R_{l_c,n, \rm ion}$, where the overall atomic index is now split into
a cell index ($l_c$) and reduced atomic index ($n$), belonging to the unit cell.
Here $S_{j,n}$ is a unitary matrix that determines the geometry of the $j$-th phonon. 
Calculations are formally performed in a box with finite length $\Lambda$, containing $N$ replicas of the unit cell. The limit 
$\Lambda \rightarrow \infty$ is eventually taken, keeping the $N/\Lambda$ unvaried. Then,
by defining $\tilde{F}_j(x_{\rm He},q)=(1/\sqrt{N}) S_{j,n}^{\dagger} e^{iq l_c L} F_{l_c,n, \rm ion}(x_{\rm He})$, 
the term $-F_{i, \rm ion} \delta R_{i,\rm ion}$
is recast in the following form: 
\begin{equation}
\label{trasfsummfr}
\sum_{q=0}^{N-1}  \tilde{F}_j(x_{\rm He},q) \delta \tilde{R}_j(-q) \,.
\end{equation}
Upon quantization of the normal vibrational modes based on quantum harmonic oscillators (QHO), Eq.~\eqref{trasfsummfr} is expressed in terms of
construction and annihilation operators ($\tilde{a}^{\dagger}_{j,q,\rm ion}$, $\tilde{a}_{j,q,\rm ion}$) such that: 
\begin{equation}
\label{trasfsummfra}
	\delta \tilde{R}_j(q) =(\tilde{a}_{j,q,\rm ion}+\tilde{a}^{\dagger}_{j,q,\rm ion})/\sqrt{2 m_{\rm C} \omega_j(q)}\,.
\end{equation}
Here $\omega_j(q)$ is the frequency of the $j$-th phonon at wavenumber $q$, and $m_{\rm C}$ is the mass of a single C atom.
Eq. \eqref{trasfsummfr} provides a coupling between He and CNT phonons, and can lead to scattering processes. 
Analogous considerations can be extended to charge displacements, hence the He-plasmon coupling term turns out to share the same 
architecture as Eq.~\eqref{trasfsummfr}, although involving the specific geometry and energy spectrum of the plasmon modes (these can
also be associated to QHO's, via analogous creation/annihilation operators).

We now estimate He-phonon scattering rates by Fermi's golden rule. 
We assume that a He atom with initial (1D) wavenumber $k_{\rm He,i}$ interacts with CNT phonons via 
Eq.~\eqref{ephcoupl}, ending up in the final wavenumber $k_{\rm He,f}$.
If the CNT initially occupies the vibrational groundstate, the transition rate is 
\begin{eqnarray}
\Gamma^{\rm ph}_{\rm i-f}=2\pi |\langle k_{\rm He,f}|-F^{\dagger}_{i,\rm ion}| k_{\rm He,i} \rangle \langle 1_{j,q} | \delta R_{i,\rm ion}| 0_{j,q} \rangle |^2
\times \nonumber \\
	 \delta(E_{\rm i,He}-E_{\rm f,He}-\omega_j(q))\,, \qquad
\label{gamma}
\end{eqnarray}
where $| 0_{j,q} \rangle$, $| 1_{j,q} \rangle$ are the groundstate and first excited state for the $j-th$ phonon at $q$. 
The delta function enforces energy conservation: in fact, the energy 
lost by He ($E_{\rm i,He}-E_{\rm f,He}$) must be converted into phonon excitation ($\omega_j(q)$). We also note that occupation of the vibrational groundstate implies a $T=0$ description. 
However, QHO excitation energies do not depend on the initial state.

We now make use of Eqs.~\eqref{trasfsummfr},\eqref{trasfsummfra}, and
consider that excitation of the $j$-th phonon with wavenumber $q$ gives 
 $\langle 1_{j,q}| \tilde{a}^{\dagger}_{j,q} | 0_{j,q} \rangle = 1$.
We also define $\Delta k_{\rm He}=k_{\rm i, He}-k_{\rm f, He}$, and note that $q$ must be compatible with the CNT unit cell. 
Recalling that $N$ unit-cell replicas are present in $\Lambda$, we facilitate normalization also assuming a finite He density, 
namely $N'$ atoms (having the same momentum for simplicity) should be present in the adopted supercell. 
We also note that ionic forces can be Fourier transformed
as: $F_{n, l_c, \rm ion}(x_{\rm He})=\frac{1}{2\pi}\int dq \tilde{f}_{n}(q) e^{iq(x_{\rm He}-l_c L)}$.  
After integration,  Eq. \eqref{gamma} finally reduces to:
\begin{eqnarray}
	\label{gammaeq}	
\Gamma^{\rm ph}_{\rm i-f}=2\pi \frac{N'}{N L^2} \bigg{\vert} \tilde{f}_{n}(\Delta k_{\rm He}) S_{n,j} \frac{1}{\sqrt{2 \omega_{j}(\Delta k_{\rm He}+mQ) m_{\rm C}}}\bigg{\vert}^2 \times \nonumber \\
	 \delta(E_{\rm i,He}-E_{\rm f,He}-\omega_j(q))\,. \qquad
\end{eqnarray}
In the $\Lambda \rightarrow \infty$ limit, the ratios $N/\Lambda$ and $N'/N$ are kept constant in order to avoid normalization issues. 
When deriving the above equation one finds that, in addition to energy conservation, {\it crystal momentum} is also conserved: 
in practice, one obtains the relation $k_{\rm i, He}=k_{\rm f, He}+q+ mQ$, where $m$ is an integer number that accounts for Umklapp processes; 
in practice, momentum is conserved up to integer multiples of the CNT lattice momentum $Q$. This property stems from the discrete
translational symmetry of the CNT. 

\begin{figure}
\includegraphics[width=8.5cm]{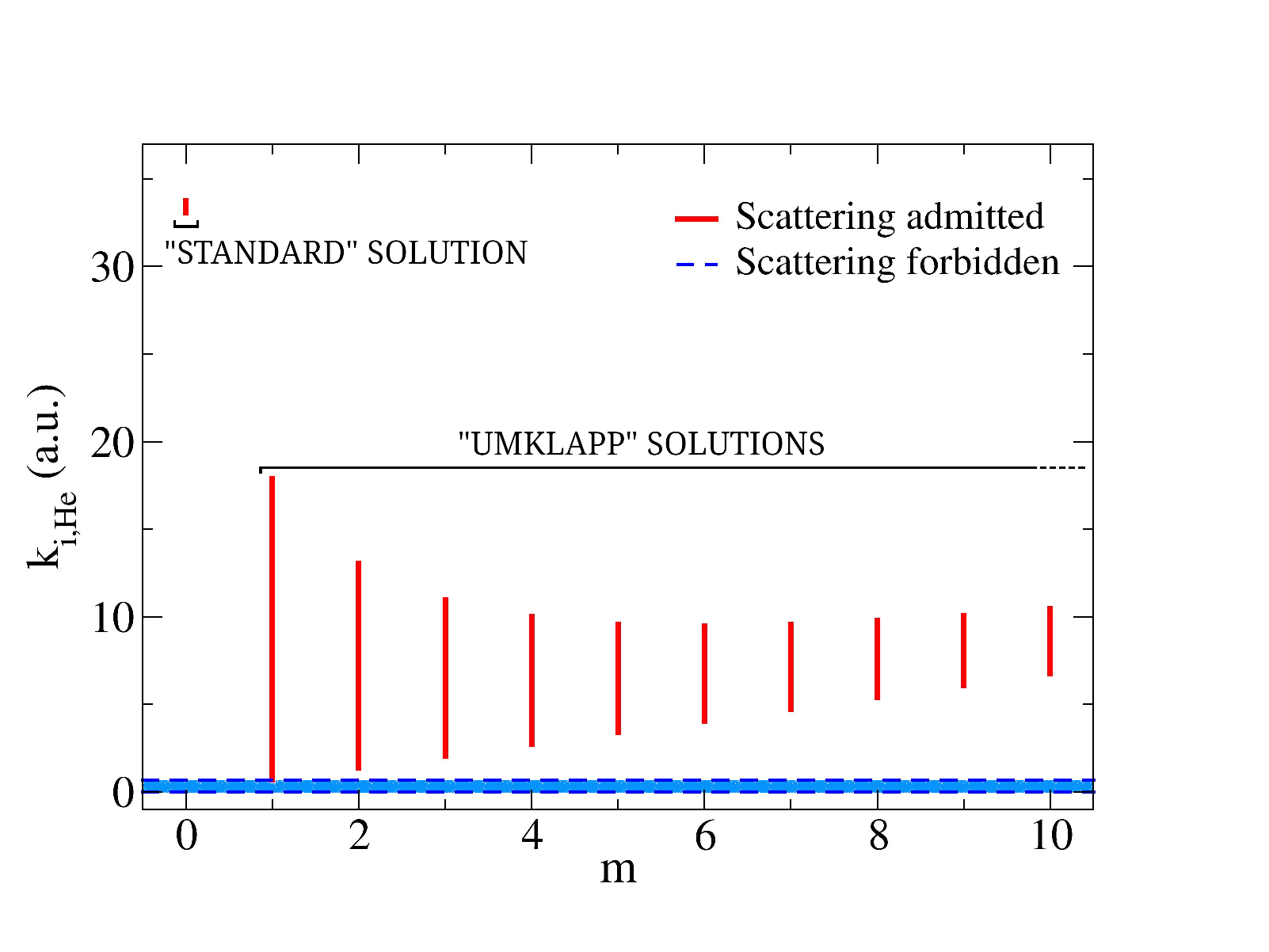}
	\caption{Graphical representation of 
	the solutions  Eq.~\eqref{soleq} relative to
	the scattering between helium an the TA phonon -- analogous solutions are found for the other excitation modes that characterize the CNT. 
	$k_{\rm i, He}$ is the wavenumber of He in the initial state (before the scattering takes place), and it is taken as positive.  Each red segment
	indicates the range of $k_{\rm i, He}$ values compatible with the He-phonon scattering at a given value of the integer parameter $m$. 
	When $m=0$ (standard solution) one has exact momentum conservation ($k_{\rm i, He}-k_{\rm f, He}=q$),
	whereas $m\neq 0$ implies occurrence of Umklapp phenomena ($k_{\rm i, He}-k_{\rm f, He}=q+mQ$). We recall that $q$ is restricted to the Brillouin zone.
	In the area delimited by dashed blue lines (colored in light blue), corresponding to $k_{\rm i, He}<Q/2$, no solution is found
	and the scattering is forbidden.
	}
\label{fig3}
\end{figure}

As in conventional superfluids, conservation of energy and momentum is traduced into a selection rule. 
At low $k_{\rm i, He}$ it is possible to adopt a free-particle dispersion for He (as justified above). 
Hence, conservation of crystal momentum and energy is expressed as: 
\begin{equation}
	\label{umkl}
\frac{k_{\rm i, He}^2-k^2_{\rm f, He}}{2m_{\rm He}}=\frac{-(q+mQ)^2+2k_{\rm i, He}(q+mQ)}{2m_{\rm He}}=\omega_j(q) \,.
\end{equation}
Eq.~\eqref{umkl} provides a generalization (due to Umklapp processes) of the familiar Landau's criterion~\cite{landau} of superfluidity, which gives the critical velocity below which the elastic collision is forbidden and the mesoscopic particle flows without friction.
According to Eq.~\eqref{umkl}, only a limited number of $k_{\rm i, He}$ values is compatible with the excitation 
of the $j$-th phonon at momentum $q$:
\begin{equation}
k_{\rm i, He}=\frac{\omega_j(q)m_{\rm He}}{q+mQ}+\frac{q+mQ}{2}\,.
\label{soleq}
\end{equation}
Recalling the linear phonon dispersion in the relevant low-momentum regime $\omega_j(q)=v_j |q|$, 
one can examine how the solutions depend on the integer Umklapp parameter $m$, with the aid of Fig.~\ref{fig3}.
At $m=0$, one has conservation of the total momentum, as in conventional superfluid regimes, and the admitted interval for the initial (positive) 
He momenta is: $k_{\rm i, He}\in[v_jm_{\rm He},v_jm_{\rm He}+Q/2]$. 
Due to large phonon velocities, very high  $k_{\rm i, He}$ is obtained. 
However, Umklapp processes significantly alter this picture, as a consequence of the CNT periodicity.
At $m=1$ the allowed momentum interval becomes $k_{\rm i, He}\in[Q/2,v_jm_{\rm He}/2+Q]$, and the lower extreme touches here the
minimum admitted value ($q$ is varied between $0$ and $Q$). 
As a consequence, no scattering is possible for $k_{\rm i, He} < Q/2$ (see Fig.~\ref{fig3}); below this threshold, friction forces are expected to vanish, 
in close analogy to standard superfluidity. 
The associated speed threshold for He superflow in the (5,5) CNT is thus $v_{\rm He}^*=Q/(2 m_{\rm He})
\sim 9.2\times10^{-5}$ a.u. (i.e. $\sim$ 200 m/s).
Conversely, friction forces are restored beyond $v_{\rm He}^*$. Taking  $k_{\rm i, He}$ above the threshold, multiple solutions
(corresponding to different $m$) can be found. However, one expects that large momentum transfer would be eventually associated
to small scattering rates, since the Fourier-transformed ionic forces $\tilde{f}_{n}$ should decay at large 
momenta.

At variance with standard superfluidity, here the critical velocity is independent from the excitation spectrum, due to Umklapp.
Coming to Plasmon excitations, analogous conclusions are drawn by approximating the spectrum as a linear function. 
Even accounting for
the logarithmic corrections to linearity expected in 1D metals, no solution is possible below $v_{\rm He}^*$; this unique critical velocity is sufficient to discriminate the {\it generalized-superfluid} regime.

By equipartion theorem, the He kinetic energy associated to the 1D critical velocity $v_{\rm He}^*$ is traduced into a temperature 
of about 20 K;
this suggests that direct injection of sufficiently slow He atoms into the CNT from a reservoir could be non-trivial. 
Nonetheless, 
major energy losses are expected when He leaves the bulk, entering the CNT edge (due to the suppression of He-He 
interactions, and collisions with the CNT edge);  He atoms in the CNT should thus be slower than expected from naive 
considerations.
In conventional superfluidity, thermal occupation of the available excitation introduces a {\it normal component} of the fluid, which
can cause scattering and finite friction. Computation of the {\it normal component} for CNT phonons and plasmon modes (see \cite{SM}) indicates that
this does not exceed $\sim0.1\%$ of the total available modes up to 300K. The stability of the {\it generalized-superfluidity}
mechanism is unparalleled, and descends from the high phonon and plamon velocities.

In summary, quantum-mechanical analysis of a He atom flowing through a sub-nm (5,5) CNT 
leads to a theoretical description which is formally similar (yet not identical) to Landau's superfluidity criterion.
The spectrum of the CNT low-lying quasi-particle excitations 
(i.e. phonon and plasmon modes) is quasi-linear with respect to momentum, although no Bose-Einstein condensation is assumed. 
This implies the existence of a critical speed $v_{\rm He}^* \simeq 200 m/s$,
below which He atoms cannot scatter against the CNT, thereby encountering no friction.
Remarkably, in the CNT $v_{\rm He}^*$ does not depend on the excitation spectrum, as a consequence of lattice periodicity
and Umklapp. Indeed, we have found that $v_{He}^* = Q/(2 m_{He})$, where
$Q=2\pi /L$ is the lattice momentum of the CNT, and $L$ its unit-cell length.
While this work specifically addresses He flow, we expect that other rare-gas atoms (or possibly other chemical moieties) could equally
move through the CNT with vanishing friction, as long as their interaction with the CNT walls is 
small enough to produce only weak geometrical perturbations. 
Analogous {\it generalized superfluidity} is expected in metallic and finite-gap CNT's
with comparable radii. Experimental validation by means of nanofluidic techniques should be viable (see \cite{SM}) due to the
simplicity and stability of the mechanism with respect to thermal excitations, and due to relatively high
$v_{He}^*$ value.

This work provides the first prediction of {\it superfluid}-like mass-transport in a standard solid system, 
and complements the ballistic electron transport already detected in CNT's.
{No ultracold gas is introduced here, spectral linearity is not strictly demanded at low momentum, and 
continuous translational invariance is not enforced.}
Extremely-high permeabilities~\cite{majumder,holt} {(3-4 orders of magnitude larger than no-slip
hydrodynamic predictions - see \cite{SM})} experimentally reported for water-flow through nanoscale
CNT’s appear qualitatively compatible with the present findings; in fact, such measurements imply drastic suppression of friction forces 
(by orders of magnitude) in the limit of small CNT radii, 
and could not be reproduced~\cite{mattia,kannam} by semi-classical models.
In spite of the higher complexity of water, a generalized superfluidity mechanism may be responsible for the observed friction suppression.
We add that enhanced nanofluidic flow was also confirmed in activated~\cite{channel} carbon channels, and high osmotic flow was
found in double-walled~\cite{osmotic} CNT's.
Extension of our quantum-mechanical theory may also interest alternative nanoscale friction~\cite{waterslide,superlubricity} 
phenomena, 
involving for instance 1D/2D heterostructures and interfaces, so that the boundary between classical and quantum-mechanical
friction mechanisms should be revisited. This work finally opens new perspectives for nanofluidics 
devices, suggesting, among others, energy-efficient quantum mechanical sieving, or non-destructive injection through cellular 
membranes.

A.A. and P.L.S. acknowledge funding from Cassa di Risparmio di Padova e Rovigo (CARIPARO) - grant EngvdW. A.A. acknowledges funding from the University 
of Padova - PARD grant. L.S. thanks Andrea Perali for useful discussions. L.S. is partially supported by the European Quantum 
Flagship Project "PASQuanS 2", by the European Union-NextGenerationEU within the National Center for HPC, Big Data and Quantum Computing 
(Project No. CN00000013, CN1 Spoke 10: “Quantum Computing”), by the BIRD Project "Ultracold atoms in curved geometries" of the
University of Padova, and by “Iniziativa Specifica Quantum” of INFN.

{A.A. conceived this work and derived the model, carried out anaytical and numerical calculations, 
wrote the article and prepared figures. P.L.S. contributed to the conceptual development of the model and to ab-initio 
calculations. L.S. introduced the interpretation of the phenomenon in terms of generalized superfluidity, contributed 
to revisions, and indicated how to perform calculations at finite-T.}

\section{Supplemental Material}

We review here in detail the approximations of the adopted model, in order to assess the
solidity of our predictions.

\subsection{Semi-local density functional theory approximation}
Semi-local exchange-correlation functionals are known~\cite{sumrule} to miss long-rage correlations, 
which implies a poor account of van der Waals forces. For this reason,
we explicitly introduced van der Waals corrections~\cite{d2} in DFT calculations.
Van der Waals corrections turn out to have negligible impact
on the corrugation of the potential experienced by He (3$\times 10^{-4}$ meV to
be compared with 0.035 meV obtained with the semi-local PBE~\cite{PBE} functional alone).
On the other hand, semi-local approximations are based on
a free electron-gas model, so that they can reliably describe
the metallic electrons present in the nanotube.
In order to exclude possible interference of the
adopted exchange-correlation approximation, we repeated DFT
calculations for the potential corrugation using a different semi-local exchange-correlation
functional, namely PBEsol~\cite{PBEsol}. We find that the potential corrugation computed
with PBEsol deviates from PBE by only $\sim8\%$, amounting to 0.032 meV.
Overall, He  remains a quasi-free particle, and no impact on the
superfluid flow is to be expected: even a variation of
the corrugation by 100$\%$ would ultimately imply no
significant change of the He spectrum reported in the article (see Fig.~2).

{\subsection{Exact exchange}
To address the possible role of exchange interactions in the corrugation of the
He-CNT potential we adopted the hybrid PBE0 functional,~\cite{pbe0} which is derived from PBE
upon inclusion of fractional exact exchange.
Given the high computational cost of the PBE0 functional~\cite{pbe0} (wich can exceed~\cite{calio} 
semi-locals by a factor of 50 or more, depending on system-size and specific implementation), 
straightforward simulation of the 8$\times$1$\times$1 supercell is very demanding.
Willing to assess the relative weight of exact
exchange with respect to other energy components already present in PBE, we thus reduced
the simulation size to the CNT unit cell, checking that the potential corrugation does
not substantially vary with respect to the cell size.
Passing from 8 to 2 unit cell replicas while keeping the reciprocal space sampling unvaried,
the PBE corrugation undergoes minor variation (from 0.035 meV to 0.029 meV). Even in the unit cell,
the corrugation remains rather stable, amounting to 0.025 meV. This demonstrates weak dependence of
He-CNT interactions with respect to the He-He separation. It is thus justified to compare
PBE and PBE0 corrugations within the unit cell. Here the PBE0 estimate is $\sim$5$\%$
larger than the corresponding PBE value. The limited role of exchange
interactions can be rationalized in terms of the small overlap between He and C electron
density tails, and is expected again to produce negligible impact on the generalized superfluidity
mechanism.}

\subsection{Ideal nanotube lattice}
Periodicity is ubiquitous in solid state physics, and does not necessarily involve
approximations. However, in the present case periodicity could be viewed as an "idealization", since:
{\it i)} experiments will necessarily deal with finite-size nanotubes;
{\it ii)} lattice defects could be present in the nanotube;
{\it iii)} deformations might be present in the structure.

Concerning point {\it i)}, as long as the finite-size nanotube is
sufficiently extended, its excitation spectrum will be distributed approximately
along the periodic bands considered in this work. 
This property can be easily deduced for charge-displacement modes, 
from existing~\cite{science,bose} literature.
Charge waves reproduce the periodicity of quasi-1D systems 
with remarkable precision, already at finite lengths of about 5-10 nm. 
No major difference is expected with vibrational modes.

A more relevant concern regards energy losses at the nanotube edges,
i.e. when a single He atom leaves or enters a reservoir. 
Hence, in order to experimentally verify the predicted
frictionless flow one should either consider long nanotubes, or
compare nanotubes with different length. Since energy losses will
be concentrated at nanotube edges, longer nanotubes must
exhibit lower effective energy-dissipation per unit-length.

Regarding points {\it ii)} and {\it iii)}, ultimate nanofabrication technologies such
as chemical vapor deposition (CVD)~\cite{cvd,cvd2}  enable realization of high-quality and virtually
defect-free nanotubes. On the other hand, nanotechnology manipulation
techniques can be exploited to fix nanotubes in the desired conformation.
A direct example is the realization of stacks of parallel nanotubes~\cite{majumder},
whereby exceedingly large geometrical distortions can be avoided.
Experimental samples can thus get surprisingly close to the
ideal situation modeled in this work.
On the other hand, we also recall that small geometrical deformations can
naturally occur due to quantum mechanical phenomenon related to phonons, and
are already accounted for in this work.

\subsection{Separation of He longitudinal and transversal motion}
The transversal motion of the He atom (orthogonally to the nanotube axis),
is controlled by the confining potential induced by interaction of He with the
nanotube walls. By sampling small He displacements from the center (within 0.3 \AA\,),
one finds that the DFT energy is well approximated by a quadratic potential, of the
form $V(r^{\rm T}_{\rm He}) \simeq V(0)+1/2 V'' (r_{\rm He}^{\rm T})^2$, where $r_{\rm He}^{\rm T}$
is the distance of He from the nanotube axis. By taking into account the atomic mass, $V''$ is
associated to a quantum oscillator frequency.
The transversal He motion can thus be described by a quantum harmonic oscillator,
whose computed frequency amounts to $\sim$8 meV. This energy scale is 
larger than the other energy scales involved in He flow: we recall that the potential corrugation
along the axis amounts to 0.035 meV, while the He kinetic energy at the critical velocity is about 0.5 meV.
One can thus factorize the longitudinal and transversal
components of the He wavefunction, and safely assume that the transversal degree of
freedom will be described by a quantum harmonic oscillator in the groundstate. 
No excitation of this transversal degree of freedom takes place at
the relevant energy scales, hence He motion can be treated as effectively 1D.

\subsection{Thermal effects}
Finite temperatures can induce thermal excitation of the plasmon and phonon modes,
which may cause scattering events. By analogy with standard superfluids, this
fraction of excited modes corresponds to the so-called {\it normal} component.
We underline that in our case no Bose-Einstein condensate is present, and no dramatic change in the
excitation spectrum should be expected up to room temperature.

We will thus simply recall that the {\it normal} component of the available excitations
can {\it move}~\cite{landau,superf} with velocity $v$ with respect to the other modes, due to momentum transfer 
induced by scattering.
The {\it normal} mass current at finite temperature $T$ is defined as:
\begin{equation}
	j_{\rm n}(T)=n_{n}(T)\, m\, v = \frac{L}{2\pi} \int_{BZ} dp\, p\, f_{\rm B}(E(p)-p v)\,,
\end{equation}
where integration is performed over the first Brillouin zone (BZ), while $n_{\rm n}$ is the {\it normal} density
component (given as the number of modes per unit cell of length $L$), $m$ is the effective mass of the excitation modes, 
and $f_{\rm B}$ is the Bose-Einstein distribution.
If small velocity $v$ is assumed, the above expression can be Taylor expanded at first order
in $v$. Hence one can identify the {\it normal} density component with
\begin{equation}
	n_{\rm n}(T) = -\frac{L}{6 k_{\rm B} \pi} \int_{BZ} dp\, \frac{p^2}{m}\, \frac{e^{E(p)/k_{\rm B}T}}{\left(e^{E(p)/k_{\rm B}T}-1 \right)^2 }\,,
\end{equation}
where $k_{\rm B}$ indicates Boltzmann's constant.
Considering phonon modes, one needs to perform a summation over all four linear branches, accounting for their energy dispersion 
(other phonons are neglected due to the finite gap). Since the nanotube is only composed of C atoms, the effective phonon mass equals the
atomic C mass.
If the phonon speeds are indicated as $v_j$, the corresponding energies are written as~\cite{dresselhaus} $E_j(p)=v_j |p|$.
When linear mode dispersion is assumed throughout the entire BZ, analytic integration of the above formula is possible, and leads to the following 
expression:
\begin{equation}
	n_{\rm n}(T) = \frac{L}{\pi m_{\rm C} v_j} \sum_{j=1}^4 \left( f_{1,j}+ f_{2,j}+ f_{3,j} \right) \,,
\end{equation}	
with
\begin{equation}
	f_{1,j}=-\frac{(\pi/L)^2}{e^{\pi v_j/Lk_{\rm B}T}} 
\end{equation}	
and
\begin{equation}	
	f_{2,j}= \frac{2 \pi k_{\rm B}T}{v_j L} ln\left( 1-e^{-\pi v_j/Lk_{\rm B}T} \right) \,,
\end{equation}	
while
\begin{equation}	
	f_{3,j}= \frac{2k_{\rm B}^2T^2}{v_j^2}\left[{\rm Li}_2(1)  - {\rm Li}_2( e^{-\pi v_j/Lk_{\rm B}T})    \right]\,.
\end{equation}	
Here ${\rm Li}_2$ is Spence's polylogarithm function, and $m_{\rm C}$ is the C atom mass. 
The {\it normal} mode component can be compared to the the {\it total} mode number
within the same unit cell (i.e. $n_{\rm tot}=4$). The ratio between {\it normal} component and {\it total} modes is plotted in Fig.~\ref{fig1} as
a function of $T$. We observe that the ratio remains very small up to room temperature (300 K). In fact, phonon speeds are high,
and excitation of high momentum modes remains difficult, even at high temperature. The small {\it normal} fraction justifies a posteriori
the adopted linear-band approximation through the entire BZ - since occupation of {\rm normal} states remains negligible.
Analogous calculations can be performed for plasmon modes, where even smaller {\it normal} fraction is found: a difference of
almost three orders of magnitude is found, due to the even higher steepness of the plasmon spectrum.
\begin{figure}
\includegraphics[width=8.5cm]{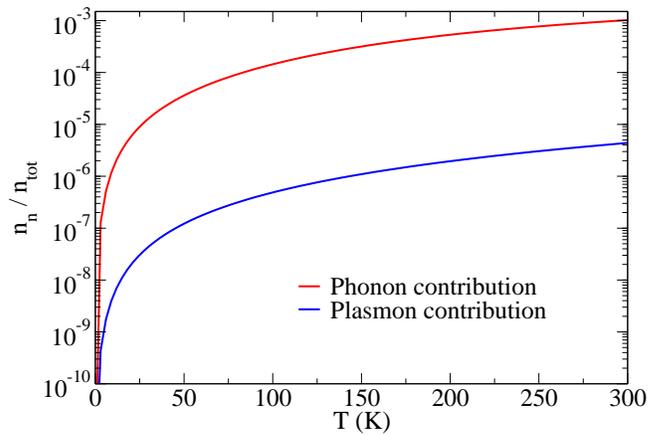}
	\caption{Ratio of the {\it normal} component vs. total available excitations as a function of the temperature $T$. Phonon and plasmon
	modes are treated separately: phonons provide the leading contribution to the {\it normal} component of the system excitations.
	}
\label{fig4}
\end{figure}

These results demonstrate that only a tiny fraction of the available modes can contribute to friction forces at temperatures that are
unparalleled by conventional superfluids. The generalized superfluid mechanism is extremely resistant to thermal effects, and 
may accordingly enable real-world technological applications.

\section{Quasi-1D model for plasmon modes}

An explicit calculation of the plasmon modes for a metallic (4,4) carbon nanotube (CNT) is provided in Ref.~\cite{mbdc}, based on a discretized model, where electron hopping is explicitly included. Given the close analogies between
the electronic structure of (4,4) and (5,5) CNT's, the properties of the plasmon modes will be transferable. In fact, the coupling between metallic (universal) and localized (system-specific) electronic modes provides modest renormalization close to $q=0$ (where $q$ is the plasmon wavenumber).

To guide intuition on plasmon modes, here we will further report a derivation based on a quasi-1D continuum model~\cite{Dobson}, which can be treated analytically, and still reproduces the main spectral features that are exploited in this work.
We consider a metallic wire, with infinite length and effective thickness $b$. In the relevant regime $qb \ll 1$ the intrawire interaction has the
form $w(q)=-2\, ln(qb)$. Atomic units are adopted, consistently with the main article.
The bare electron-density susceptibility is 
\begin{equation}
\chi^0(q)=(N_0 q^2/\omega^2)\,,
\end{equation}
where $N_0$ is the number of metallic electrons per unit length and $\omega$ is the frequency.
Within the random phase approximation, the poles of the interacting response function satisfy the equation
\begin{equation}
	\chi^0(q) w(q) = 1\,,
\end{equation}
which leads to the following solution for the the plasmon frequency:
\begin{equation}
	\omega = |q| \sqrt{2 N_0 |ln(qb)|}\,.
\end{equation}
According to this expression, one has a linear dispersion, corrected by a logarithmic factor. The plasmon frequency
vanishes in the $q\rightarrow 0$ limit, so that the logarithmic correction has no effect on the critical velocity (see Eq.~8 in the main article).

\section{Connection to experiment and nanotube features}
{The detection of enhanced water flow can be experimentally accomplished by permeability
measurements. When a pressure difference $\Delta P$ is applied between two water
reservoirs connected by a CNT, one detects a net volumetric flux $Q$. The volumetric
flux relates to the permeability $\kappa$ through the following relation:
\begin{equation}
Q = \frac{\kappa \, \Delta P S}{L_{\rm tot}} \,,
\end{equation}
where $S$ and $L_{\rm tot}$ are the CNT transversal area and total length, respectively.
Classical macroscopic hydrodynamics (no-slip Haagen-Poiseuille theory~\cite{classic}), predicts
the following permeability:
\begin{equation}
\kappa_{\rm HP}=\frac{R^2}{8\mu}\,,
\end{equation}
where $R$ is the CNT radius and $\mu$ is the dynamic viscosity of water.
However, experimental permeabilities~\cite{holt} can easily exceed this classical estimate by 3-4 orders
of magnitude when CNT radii approach the nanometer scale. This implies a breakdown of classical fluid
dynamics, which is associated to extremely low friction, and slippage of water molecules with respect
to the CNT walls. Extension of the aforementioned approach could also be exploited for the experimental
detection of generalized He superfluid flow, given the direct analogy between the two setups.}

The main prerequisite for the onset of frictionless He flow
is the respect of the maximum atomic speed (which should be smaller
than the critical velocity). If He is extracted from a reservoir at
finite temperature, it is probable that scatterings at the nanotube
edges can significantly reduce the thermal velocity of the atom. 
In the same breath, when
a single He atom is extracted from the reservoir, the former interaction with the
surrounding He gas implies additional energy loss. It is thus plausible that
He gases well above 20 K can be used as a reservoir, still respecting
the critical velocity limit.

Other experimental issues regard the quality of the nanotube (presence of
defects, or distortions) which should be kept under control: defects may
act as scattering sources. A possible strategy to control mechanical
distortions is the realization of porous membranes composed by a stack of
parallely oriented nanotubes: the bending of stacked nanotubes is expected to be unfavored.

The choice of a (5,5) carbon nanotube is mainly due to two reasons:
{\it i)} The diameter of this nanotube roughly matches the sum of the C and He van der Waals radii, so
that only a single He atom can fit in the nanotube section. At the same time, the
Pauli repulsion between He electronic cloud and nanotube walls remains weak. Under these
conditions, the He atom undergoes effective 1D motion, while the
He-nanotube interaction potential exhibits very small oscillations.
{\it ii)} The (5,5) nanotube is metallic, so that both phonon and plasmon modes can be excited
by the travelling He atom. This implies a richer physical model. In a finite-gap nanotube with comparable radius
one only expects phonon modes to be relevant. Hence, the frictionless flow should be 
be preserved, although somewhat larger potential oscillations are expected due to charge localization at C-C bonds.

In summary, the physical effect described in this work should reasonably hold both in metallic and finite-gap nanotubes 
with comparable radii.

When the radius is too small, He atoms  undergo large Pauli Repulsion, and their spectrum can
significantly deviate from the free-particle parabolic dispersion seen here.
Conversely, when the radius is much larger, He may stick to the nanotube wall, staying away from the
center. The nanotube walls become more deformable, the motion can deviate from the considered 1D trajectories, and several He atoms 
can  pass at the same time through the nanotube section.
We also expect that in large nanotubes the flowing He atoms could thermalize with the reservoir, so
that the critical velocity threshold may be harder to respect.

While no direct experimental validation exists yet for the predicted frictionless
He flow, a number of experiments have been conducted for
water flow in carbon nanotubes (see for instance Refs.~\cite{holt,majumder}).
These experiments consistently indicate drastic suppression of friction forces when
the nanotube radius approaches the nm scale. Moreover, permeabilities steeply
increase when the radius is further decreased.

All attempts to model water flow within semi-classical theories
have failed~\cite{mattia,kannam} in reproducing the extremely high permeabilities measured by
experiments. The underlying mechanism should accordingly have a truly
quantum-mechanical character.
The sought-after quantum-mechanical effect should persist in the limit of very narrow
nanotubes, and should cause drastic suppression of water-nanotube scattering.

The generalized superfluidity mechanism could provide, by extension~\cite{palermo} of our work,
a clean physical interpretation for the above experimental observations.
Indeed, some more issues should be addressed before generalized superfluid flow of water is
confirmed: water implies complications with respect to He, due to the
internal structure and the ability to form relatively strong intermolecular
bonds.

He and other noble-gas atoms enable simplification of the problem
both from the theoretical and experimental point of view, thereby
providing a more convenient platform for test and characterization
of the generalized superfluid phenomenon.

{\section{Remarks}
Landau's account of standard superfluidity is deeply rooted in translationally-invariant ultracold gases. 
In fact, Landau's theory involves as a key step the derivation of the quasiparticle excitation modes of 
an ultracold gas. In particular, a linear phonon spectrum is obtained at low momenta exploiting 
continuum hydrodynamics equations. The collective quasi-particle modes of the ultracold gas are 
fundamental here, since they are identified as the available excitation modes of the system.
Energy-momentum conservation directly follows as a standard condition that should
be respected in scattering events. Accordingly, Landau's theory is broader than just spectral 
incompatibility alone.
\\
\indent
In this letter we did not assume the presence of any ultracold gas, we did not introduce Bogoliubov excitations, 
we did not exploit hydrodynamics equations, and we neither assumed continuous translational invariance 
(the CNT is characterized by a discrete lattice). We derived instead a fully quantum mechanical model 
for the scattering between He and CNT plasmon and phonon modes, based on first-principle calculations 
and Fermi's golden rule. The excitation modes of the CNT are not the same as in conventional superfluids.
While phonons exhibit linear spectrum at vanishing momentum, plasmons are always characterized by 
logarithmic corrections, hence they qualitatively differ from Bogoliubov excitations. As predicted by Fermi's 
golden rule, He-CNT scattering necessarily involves two-fold conservation of energy and quasi-momentum. 
But quasi-momentum also differs from the standard linear momentum conserved in Landau's theory due to the 
discrete CNT lattice. For this reason, our critical velocity contrasts with Landau's prediction, and does 
not depend on the excitation spectrum.
These discrepancies ultimately imply unparalleled robustness of the generalized superfluidity mechanism 
with respect to thermal excitations: the {\it normal} component of available CNT excitation modes exhibits 
minimal relative weight up to room temperature - again at stark variance with conventional superfluids.
\\
\indent
Considering that CNT's are solid nanostructures that have little to do with conventional superfluids, 
the emergence of a generalized superfluid mechanism which can persist even at high temperatures is 
undoubtedly surprising.
Last but not least, the application spectrum of generalized superfluidity markedly differs from conventional 
superfluidity: the former may encompass gas filtration by CNT stacks at low energy cost, non-destructive 
injection through cellular membranes via CNT's, or realization of highly efficient nanofluidic circuits. 
In addition, surprisingly-low friction forces may eventually interest other quasi-1D or 2D nanointerfaces, 
thereby implying novel superlubricity mechanisms.}


\end{document}